\documentclass{article}
\def\1ad{\mbox{\normalsize $^1$}}
\def\2ad{\mbox{\normalsize $^2$}}
\def\3ad{\mbox{\normalsize $^3$}}
\def\4ad{\mbox{\normalsize $^4$}}
\def\5ad{\mbox{\normalsize $^5$}}
\def\6ad{\mbox{\normalsize $^6$}}
\def\7ad{\mbox{\normalsize $^7$}}
\def\8ad{\mbox{\normalsize $^8$}}
\def\makefront{
\hfill HU-EP-05/07
\vspace*{1cm}
\begin{center}
\def\sp{
\renewcommand{\thefootnote}{\fnsymbol{footnote}}
\footnote[1]{corresponding author~~E-mail: \email_speaker}
\renewcommand{\thefootnote}{\arabic{footnote}}
}
\def\newtitleline{\\ \vskip 5pt}
{\Large\bf\titleline}\\
\vskip 1truecm
{\large\bf\authors}\\
\vskip 5truemm
\addresses
\end{center}
\vskip 1truecm
{\bf Abstract:}
\abstracttext
\vskip 1truecm
}

\usepackage{latexsym,amsfonts}
\newcommand{\rr}{\mathbb{R}}
\newcommand{\hh}{\mathbb{H}}
\def\beq{\begin{equation}}                     %
\def\eeq{\end{equation}}                       %
\def\bea{\begin{eqnarray}}                     
\def\eea{\end{eqnarray}}                       
\begin {document}
\def\titleline{
%
%
%
On particle dynamics in $AdS_{N+1}$ space-time                       
}
\def\email_speaker{
{\tt
%
%
jorj@rmi.acnet.ge            
}}
\def\authors{
%
%
%
%
%
Harald Dorn\1ad , George Jorjadze\1ad \2ad
}
\def\addresses{
%
%
%
%
\1ad
Institut f\"ur Physik der
Humboldt-Universit\"at zu Berlin,\\
Newtonstra{\ss}e 15, D-12489 Berlin, Germany\\[1mm]
\2ad Razmadze Mathematical Institute,\\ M.Aleksidze 1, GE-0193
Tbilisi, Georgia
}
\def\abstracttext{
%
%
We summarize part of a systematic study of particle dynamics
on $AdS_{N+1}$ space-time based on Hamiltonian methods. New explicit
UIR's of $SO(2,N)$, defined on certain spaces of holomorphic functions,
are constructed. The connection to some field theoretic
results, including the construction of propagators, is discussed.
}
\large
\makefront
\section{Introduction}
The concept of quantized particles in Minkowski space is intimately
connected with the unitary irreducible representations (UIR's) of
its isometry group, the Poincare group. In generic curved
spacetime manifolds, due to the lack of enough isometries, the
concept of quantized particle becomes dubiously. The description
of matter is more appropriate in terms of quantized fields,
although part of the difficulties then comes back via ambiguities
in defining vacuum states.

In $AdS$, as a constant curvature background, one has an isometry group of the
same dimensionality as in Minkowski space and particle quantization along the
standard Hamiltonian techniques is possible. This has been done for lower
dimensional cases \cite{J,Ful,Luc,Siop}. Our present contribution will summarize some aspects of our
study of both the classical and quantum particle dynamics on $AdS_{N+1}$ for
general $N$. With this study we will find new explicit UIR's of
$SO(2,N)$. Another motivation for our investigation comes from possible
interrelations with the corresponding quantum field theories on $AdS$, which
play a crucial role in the $AdS$/CFT correspondence, see e.g. \cite{ADSCFT,HF}.

In this note we do not refer to the reach literature on $SO(2,N)$
representations in detail, leaving this account for an extended  work in
progress \cite{DJ}.

The $AdS_{N+1}$ space can be related to $\hh_N^R$, the hyperboloid
$ X_{0'}^2+X_0^2-X_nX_n=R^2$ embedded in $\rr_N^2$ with metric
tensor $\,G_{AB}=\mbox{diag}(+,+,-,...,-)$. As global coordinates
we use $x_0=\theta,~x_n,~~(n=1,...,N)$
\begin{eqnarray}\label{x}
X_0=r\,\cos\theta~,~~~~X_{0'}=r\,\sin\theta~,~~~~X_n=x_n~,~~~~
\mbox{with}~~~~r=\sqrt{R^2+x_nx_n}~.
\end{eqnarray}
The cyclic coordinate $\theta\in{\cal S}^1$ is
identified with time. Obviously, $SO(2,N)$ is the isometry group of
the obtained space-time. Unwrapping the time coordinate
$\theta\in \rr^1$, one gets the universal covering of $\hh_N^R$,
and $AdS_{N+1}$ is usually associated with it.
\section{Classical dynamics}
The dynamics of a particle with mass $\,m\,$ moving on
$\hh_N^R$  is described by the action
\begin{equation}\label{S}
S=-\int d\tau \Big (\frac{\dot X^A\dot X_A}{2e}+\frac{em^2}{2}+
\frac{\mu}{2}(X^AX_A-R^2)\Big )~,
\end{equation}
where $e$ and $\mu$ are Lagrange multipliers and
$\tau$ is an evolution parameter.
To fix the time direction on $\hh_N^R$, we make the $SO(2,N)$ invariant
choice \cite{DJ} $\dot\theta>0$, which via (\ref{x}) is equivalent to
\begin{equation}\label{dot-t}
 {X_0\dot X_{0'}-X_{0'}\dot X_0}>0~.
\end{equation}
We also assume $e>0$, to have a positive kinetic term for the spacelike
 coordinates.

The $SO(2,N)$ symmetry of (\ref{S}) provides
the Noether dynamical integrals
\begin{equation}\label{M_AB}
J_{AB} =P_A\,X_B -P_B\,X_A ~,
\end{equation}
where $P_A$ are the canonical momenta $P_A= -{\dot X_A}/e$.
The generators $J_{0n}$ and $J_{0'n}$
are related to the non compact $SO(2,N)$ transformations, while
$J_{00'}$ and $J_{mn}$ to the compact ones,
which form the subgroup $SO(2)\times SO(N)$. We will use the notations
$J_{0n}=K_n$, $J_{0'n}=L_n$ and $J_{00'}=E$.
Since $\theta$ is the time coordinate, $E$ is identified
with the particle energy,
and due to (\ref{dot-t}) it is positive.

The dynamical integrals (\ref{M_AB}) allow to represent the set of
all trajectories geometrically without solving the dynamical
equations. From (\ref{M_AB}) we find N equations
\begin{equation}\label{traj}
E\,X_n=K_{n}\,X_{0'}-L_{n}\,X_0~,
\end{equation}
which define a 2-dimensional plane in the embedding space $\rr_N^2$,
containing the origin. Its intersection with the hyperboloid is
a particle trajectory.

For further calculations it is convenient to introduce the
complex variables
\begin{equation}\label{z_n}
z_n=L_{n}-iK_{n}~,~~~~~~~~~~z_n^*=L_{n}+iK_{n}~.
\end{equation}
Then, from (\ref{x}) and (\ref{traj}) we obtain
\begin{eqnarray}\label{traj1}
&&X_0=r(\theta)\cos\theta~,~~~X_{0'}=r(\theta)\sin\theta~,
~~~X_n=-\frac{r(\theta)}{2E}\left(z_n^*\,e^{i\theta}+
z_n\,e^{-i\theta}\right),~~ \\
\nonumber
&\mbox{where}~~&
r(\theta)=\frac{2ER}{\sqrt{4E^2-2\lambda^2-2\rho^2\cos(2\theta+\beta)}}~,\hfill
~\\
\label{z^2}
&&\lambda^2=z^*_nz_n~,~~~~~~~\rho^2\,e^{-i\beta}=z_nz_n~,~~~~~~~~~
\rho^2\,e^{i\beta}={z^*_nz^*_n}~.~~~~~~~~~~~~~~~~
\end{eqnarray}

The action (\ref{S}) is gauge invariant under the reparameterizations
$\tau\rightarrow f(\tau)$ combined with $\mu \rightarrow \mu\,/ f'$,
$e \rightarrow e/ f'$. This gauge symmetry, as usual, leads to
dynamical constraints. Applying Dirac's procedure, we find three
constraints
\begin{equation}\label{Phi=0}
X^AX_A-R^2=0~,~~~~~~~~P_AP^A-m^2=0~,~~~~~~~~~
P_A\, X^A=0~,
\end{equation}
which fix the quadratic Casimir number
of the symmetry group
\begin{equation}\label{C}
C=\frac{1}{2}\, J_{AB}J^{AB}=m^2R^2~.
\end{equation}
This equation can be rewritten as
\begin{equation}\label{C1}
E^2+J^2=\lambda^2+\alpha^2~,~~~~~~~\mbox{with}~~~~
J^2=\frac{1}{2}\, J_{mn}J_{mn}~~~~~~~\mbox{and}~~~~~~~\alpha=mR~.
\end{equation}
Another set of quadratic
relations follows from (\ref{M_AB}) as identities
in the $(P, X)$ variables
\begin{equation}\label{MM=MM}
J_{AB}\,J_{A'B'}= J_{AA'}\,J_{BB'}-J_{AB'}\,J_{BA'}~.
\end{equation}
These equations are nontrivial in terms of the symmetry generators,
if all indices are different. Taking $A=0$, $B=0'$, $A'=m$
and $B'=n$ $(m\neq n)$
we obtain
\begin{equation}\label{EM=zz}
2iE\,J_{mn}=z_m^*z_n-z_n^*z_m~,
\end{equation}
which provides
\begin{equation}\label{EM=lr}
4E^2J^2=\lambda^4-\rho^4~.
\end{equation}
By eqs (\ref{C1}) and (\ref{EM=lr}) $E^2$ and $J^2$ are roots
of the quadratic equation
\begin{equation}\label{E=?}
4x^2-4(\lambda^2+\alpha^2)x+\lambda^4-\rho^4=0~.
\end{equation}
We choose
\begin{eqnarray}\label{E=}
E^2=\frac{1}{2}\,\left(\lambda^2+\alpha^2+
\sqrt{\alpha^4+2\alpha^2\lambda^2+\rho^4}\,\right),\\
\label{M^2=} J^2=\frac{1}{2}\,\left(\lambda^2+\alpha^2-
\sqrt{\alpha^4+2\alpha^2\lambda^2+\rho^4}\,\right).\,
\end{eqnarray}
The other choice would be unphysical,
since with $E^2$ chosen as the small root of (\ref{E=?}), $r(\theta)$ in
(\ref{traj1}) would become imaginary. According to (\ref{E=}) $\alpha$ is
the lowest value of energy. Using in addition (\ref{M^2=}), one can show
that $E\geq J+\alpha$.
\\

We now study Hamiltonian reduction.
The generators $E$ and $M_{mn}$
are functions of ($z_n$, $z_n^*$) via (\ref{E=}) and
(\ref{EM=zz}). Therefore,
($z_n$, $z_n^*$) $\sim$ ($K_n$, $L_n$) are global coordinates on
the space of dynamical integrals.
There are no restrictions on these coordinates
for $\alpha>0$ and they cover all $\rr^{2N}$, but for the massless
case one has to remove the submanifold $z_nz_n=0$.

Eq. (\ref{traj1}) can be considered as a parameterization of
$X_A$ by the dynamical integrals
and the time coordinate $\theta$.
The canonical momenta $P_A$ are parameterized
in a similar way. These parameterizations together
define the gauge orbits on the $2N+1$ dimensional
surface constrained by (\ref{Phi=0}),
and $\theta$ is the parameter along the orbits.
The physical phase space  $\Gamma_{ph}$ is the set of these
orbits, it is $2N$-dimensional and is given
by the gauge invariant variables $z_n$, $z_n^*$.
Thus, $\Gamma_{ph}$ is identified with a $2N$-dimensional
space of dynamical integrals.

Our aim is to describe the Poisson bracket structure on  $\Gamma_{ph}$
and to calculate the corresponding reduced
symplectic form $\omega$. Starting with the canonical
$\{P_A,X^B\}=\delta_A\,^B$ the Poisson bracket algebra of the generators
(\ref{M_AB}) turns out to be $so(2,N)$
\begin{equation}\label{PB_M}
\{J_{AB},\,J_{A',B'}\}=
G_{AA'}J_{BB'}+G_{BB'}J_{AA'}-
G_{AB'}J_{BA'}-G_{BA'}J_{AB'}~.
\end{equation}
Since the generators are gauge invariant, their Poisson bracket
algebra is the same after the reduction to $\Gamma_{ph}$ and we have
\begin{eqnarray}\label{PB_z}
&&\{z_m^*,\,z_n\}=2J_{mn}-2i\delta_{mn}\,E~,\\
&&\{E,\,z_n\}=-iz_n~,~~~~~~~~~~
\{J_{lm},\,z_n\}=z_l\,\delta_{mn}-z_m\,\delta_{ln}~,\\ \label{PB_z=0}
&&\{z_m,\,z_n\}=0=\{z_m^*,\,z_n^*\}~,~~~~~~\{E,\,J_{mn}\}=0~.
\end{eqnarray}
These Poisson brackets are essentially
non-linear in terms of the independent variables $z_n$, $z_n^*$
and their quantum realization becomes problematic.
Below we apply the method of geometric quantization,
which is based on the symplectic structure of the classical system.
The symplectic form $\omega$ can be obtained
as a reduction of the canonical
form $dP_A\wedge dX^A$. To simplify this procedure,
we use the relation valid on the constraint surface (\ref{Phi=0})
\begin{equation}\label{omega_0}
dP_A\wedge dX^A=\frac{1}{2\alpha^2}J_{AB}dJ^{AC}\wedge dJ^B\,_C~.
\end{equation}
Then, parameterizing all generators by
$z_n$ and $z_n^*$, a straightforward calculation gives
\begin{equation}\label{omega}
\omega=\omega_{mn}\frac{dz_m\wedge dz_n^*}{2iE}~,
\end{equation}
where
\begin{eqnarray}\label{omega_mn}
\omega_{mn}=\delta_{mn}-\frac{{z^*}^2}{4E^2(E^2-J^2)}\, z_m z_n-
\frac{{z}^2}{4E^2(E^2-J^2)} \, z_m^* z^*_n \nonumber \\
+\frac{z^*z+2E^2 }{4E^2(E^2-J^2)}z_m z^*_n
+\frac{z^*z-2E^2 }{4E^2(E^2-J^2)}z_m^* z_n~,
\end{eqnarray}
and use has been made of (\ref{E=}) and (\ref{M^2=}).

Since the phase space for $\alpha>0$ has the structure of $\rr^{2N}$
the symplectic form (\ref{omega}) is exact $\omega =d\Omega$.
The integration of the 2-form (\ref{omega}) yields
\begin{equation}\label{Theta}
\Omega=\frac{1}{4iE}\left(z_n+\frac{{z}^2}{F^2}\,z^*_n\right)\,dz^*_n
-\frac{1}{4iE}\left(z^*_n+\frac{{z^*}^2}{F^2}\,z_n\right)\,dz_n~,
\end{equation}
with
\begin{equation}\label{F=}
F=\sqrt{(E+\alpha)^2-J^2}~.
\end{equation}
\section{Quantization}
The quantum theory of particle dynamics on $AdS$ space can be
constructed quantizing the classical system given on
$\Gamma_{ph}$. A consistent quantization should provide an unitary
irreducible representation of the symmetry group in some Hilbert
space. The cases $N=1$ and $N=2$ have been discussed in several
papers \cite{J}-\cite{Siop}. $N=2$ splits in two $N=1$ cases. For
$N=1$ one usually uses the (Bargman) Hilbert space \cite{K} formed
by the holomorphic functions $\psi(\zeta)$ inside the unit disk
$|\zeta|<1$  with the scalar product
\begin{equation}\label{sc-pr}
\langle\psi_1|\psi_2\rangle =
\int_{\zeta|<1}d\zeta\,d\zeta^*\,(1-|\zeta|^2)^{2\alpha-2}\,\,
\psi_1^*(\zeta)\,\psi_2(\zeta)~.
\end{equation}

The dynamical integrals (\ref{M_AB}) are transformed by
the co-adjoint representation of $SO(2,N)$ group
$J_{AB}\rightarrow \Lambda_A\,^{A'}\Lambda_B\,^{B'}J_{A'B'}$,
 and $\Gamma_{ph}$
is the orbit of the point $z_n=0$, $J_{mn}=0$ and
$E=\alpha$. Therefore, it is natural to apply the method of
geometric quantization \cite{W}.

At the first stage of this approach we construct pre-quantization
operators. They are given by the map $f\mapsto \hat O_f$
from observables $f$, which are functions on $\Gamma_{ph}$,
to the operators $\hat O_f$ acting
on the Hilbert space ${\cal L}^2(\Gamma_{ph})$
\begin{equation}\label{O_f}
\hat O_f=f-\Omega(V_f)-iV_f~.
\end{equation}
Here $V_f$ is the Hamiltonian vector field
\begin{equation}\label{V_f}
V_f=\{f,z_n\}\,\partial_{z_n}+ \{f,z^*_n\}\,\partial_{z^*_n}~,
\end{equation}
and $\Omega(V_f)$ is the value of the 1-form (\ref{Theta}) on $V_f$.
 $\Omega(V_f)$ is calculated in a standard
way $\Omega(V_f)=\Omega_{z_n}\{f,z_n\}
+\Omega_{z^*_n}\{f,z^*_n\}$, where $\Omega_{z_n}$ and
$\Omega_{z^*_n}$ are the coefficients in (\ref{Theta}) of the
differentials $dz_n$ and $dz^*_n$,   respectively. The operators
(\ref{O_f}) act on wave functions $\Psi(z,z^*)$ and they are
hermitian with respect to the scalar product based on the
Liouville measure on $\Gamma_{ph}$. The Hamiltonian vector fields
of the symmetry generators are obtained from (\ref{V_f}) and after
calculation of the values of $\Omega$ on these fields we find the
pre-quantization operators
\begin{eqnarray}\label{O_E}
&&\hat O_E=\alpha - z_n\partial_{z_n}+z^*_n\partial_{z^*_n}~,\\ \label{O_J}
&&\hat O_{J_{mn}}=i~(z^*_n\partial_{z^*_m}-z^*_m\partial_{z^*_n})+
i~(z_n\partial_{z_m}-z_m\partial_{z_n})~,\\ \label{O_z}
&&\hat O_{z_n}=\frac{\alpha}{2E}\Big (z_n+\frac{z^2}{F^2}\,z^*_n\Big )+
2(\delta_{nm}\,E-iJ_{nm})~\partial_{z^*_m}~,\\ \label{O_z^*}
&&\hat O_{z^*_n}=\frac{\alpha}{2E}\Big (z^*_n+\frac{{z^*}^2}{F^2}\,z_n\Big )-
2(\delta_{nm}\,E+iJ_{nm})~\partial_{z_m}~.
\end{eqnarray}
The operators (\ref{O_E})-(\ref{O_z^*})
give an unitary representation of the $so(2,N)$ algebra,
but it is not a representation we are looking for, since it
is reducible and the spectrum of $\hat O_E$
is not positive.
 The representation becomes
irreducible on a subspace of ${\cal L}^2(\Gamma_{ph})$ which is defined by
a choice of polarization \cite{W}.
This subspace we will construct in a more physical way, rather than via a
choice of polarization.

Let us define the vacuum
as a lowest energy state, which is invariant under $SO(N)\times SO(2)$, i.e.
\begin{equation}\label{Vac}
\hat O_{J_{mn}}\,\Psi_0=0~,~~~~~~~~~\hat O_E\, \Psi_0=\alpha
\Psi_0~,~~~~~~~~~ \hat O_{z_{n}}\,\Psi_0=0~.
\end{equation}
By the first two equations $\Psi_0(z,z^*)$ depends only
on the scalar quantities $\lambda ^2$ and $\rho ^2$ from (\ref{z^2}), and
then, due to the third
equation we find
\begin{equation}\label{Vac1}
\Psi_0={c_N}\,F^{-\alpha}~,
\end{equation}
where $c_N$ is a normalization constant
and the function $F$ is given by (\ref{F=}).
Our representation subspace  we will form by linear combinations of the states
$\Psi =(\hat O_{z^*_{1}})^{{n_1}}(\hat O_{z^*_{2}})^{{n_2}}...\,\Psi_0.$
The action of $\hat O_{z_n}$ on the vacuum state is given by
\begin{equation}\label{O_zpsi}
 \hat O_{z_{n}^*}\,\Psi_0= 2\alpha\zeta_n\,\Psi_0~,
\end{equation}
where $\zeta_n$ are complex variables
\begin{equation}\label{zeta}
\zeta_n=\frac{1}{2E}\left(z^*_n+\frac{{z^*}^2}{F^2}\,z_n\right).
\end{equation}
On the level of $SO(N)$ scalars the last definition implies
\begin{eqnarray}\label{zeta^2}
\zeta^2=\frac{{z^*}^2}{F^2}~, ~~~~~
\zeta^*\zeta=\frac{E^2-J^2-\alpha^2}{(E+\alpha)^2-J^2}~,
~~~~~~{\zeta^*}^2\zeta^2=\frac{(E-\alpha)^2-J^2}{(E+\alpha)^2-J^2}~,
\end{eqnarray}
with the  notations:
$\zeta^2=\zeta_n\zeta_n\,$,
$\,{\zeta^*}^2=\zeta^*_n\zeta^*_n\,$, $\,\zeta^*\zeta=\zeta^*_n\zeta_n\,$
$\,{z^*}^2=z^*_nz^*_n$.
This can be used to express $E$ and $F^2$ in terms of the $\zeta$'s
\begin{eqnarray}\label{E,F=zeta}
F^2=\frac{4\alpha^2}{{1-2\zeta^*\zeta+{\zeta^*}^2\,\zeta^2}} ~,
~~~~~
E=\alpha\,\frac{1-{\zeta^*}^2\,\zeta^2}{1-2\zeta^*\zeta+{\zeta^*}^2\,\zeta^2}~,
\end{eqnarray}
and to
finally arrive at the inversion of  (\ref{zeta})
\begin{eqnarray}\label{z_zeta}
z^*_n=\frac{2\alpha\,(\zeta_n-\zeta^2\,\zeta^*_n)}
{1-2\zeta^*\zeta+{\zeta^*}^2\,\zeta^2}~.
\end{eqnarray}
Thus, $\zeta$, $\zeta^*$ are global coordinates on $\Gamma_{ph}$, and due to
(\ref{zeta^2}) and (\ref{E,F=zeta})
they are constrained to the domain
\begin{equation}\label{domain}
\zeta^*\zeta <1~,~~~~~~~1-2\zeta^*\zeta+{\zeta^*}^2\,\zeta^2 > 0~.
\end{equation}
The $\zeta_n$ variables have the remarkable property  that the action of
the Hamiltonian vector fields of the symmetry generators
(\ref{V_f}) on $\zeta_n$'s
is expressed in terms of the $\zeta_n$'s alone
\begin{eqnarray}\label{PB:E,zeta}
&&V_E(\zeta_n)=i\zeta_n~,~~~~~~~~~
V_{J_{lm}}(\zeta_n)=\delta_{mn}\,\zeta_l -\delta_{ln}\,\zeta_m~,\\
&&V_{z_m}(\zeta_n)=i\delta_{mn}~,~~~~~~~\,
V_{z^*_m}(\zeta_n)=2i\zeta_m\zeta_n-i\delta_{mn}\,\zeta^2~.\nonumber
\end{eqnarray}
These equations provide the following
structure of the states
\begin{equation}\label{psi_}
\Psi=\psi(\zeta_1,...,\zeta_N)\,\Psi_0~,
\end{equation}
which is invariant under the action of the
symmetry generators (\ref{O_E})-(\ref{O_z^*}).
This yields an irreducible
representation of the $SO(2,N)$ group on the space  ${\cal H}_{ph}$
of holomorphic functions $~\psi(\zeta_1,...,\zeta_N)$.
Recalculating the operators (\ref{O_E})-(\ref{O_z^*}) on  ${\cal H}_{ph}$
we find the following representation of the $so(2,N)$ algebra
\begin{eqnarray}\label{repr}
\hat E~=&\alpha + \zeta_n\,\partial_{\zeta_n}~,~~~~~~~~~~~~~~~~~~~~~~~~~~~~~
\hat J_{mn}~=&i(\zeta_n\,\partial_{\zeta_m}-\zeta_m\,\partial_{\zeta_n})~,\\
\hat z^*_n~=&2\alpha\zeta_n+(2\zeta_n\zeta_m-\zeta^2\,\delta_{mn})\,
\partial_{\zeta_m}~,~~~~~~~~\hat z_n~=&\partial_{\zeta_n}~.\nonumber
\end{eqnarray}
The energy spectrum on ${\cal H}_{ph}$ is positive
$\langle\hat E\rangle\geq \alpha$.

Recalculating the Casimir number of this representation one finds a quantum deformation relative
to the classical $C=\alpha ^2~:$
\begin{equation}\label{C^}
C=\hat E^2+\frac{1}{2}\, \hat J_{mn}\hat J_{mn}-\frac{1}{2}
\,(\hat z^*\hat z +\hat z\hat z^*)=
\alpha(\alpha-N)~.
\end{equation}
Starting with the Liouville measure and using (\ref{psi_}) and
(\ref{Vac1}) one finds for the scalar product on ${\cal H}_{ph}$
\begin{equation}\label{psi|psi}
\langle\psi_1|\psi_2\rangle =  \int d^{N}\zeta\, d^{N}\zeta^*
\,(1-2\zeta^*\zeta+{\zeta^*}^2\zeta^2)^{\alpha-N}\,\,
\psi_1^*(\zeta)\,\psi_2(\zeta)~.
\end{equation}
For $N=1$ this result coincides with (\ref{sc-pr}). The
integration measure in the scalar product for $N=1$ is regular for
$\alpha>1/2$, however, the regularity of the integration measure
for $N\geq 2$ requires $\alpha>N-1$. Thus, for $N\geq 2$ and
$\alpha>N-1$ eqs. (\ref{repr}) and (\ref{psi|psi}) define UIR's of
the $SO(2,N)$ group with the Casimir number $C=\alpha(\alpha-N)$.

In this deformed relation $C$ has to be interpreted
as the squared mass of the $quantum$ particle (multiplied by $R^2$). The
parameter $\alpha$,
as shown above, is the lowest value of the energy within the
representation. The well-known unitarity bound for UIR's of $SO(2,N)$, see
e.g. \cite{ADSCFT,HF}, given by
$\alpha\geq N/2-1$, is of course respected by our representations. However, it
remains to be clarified why these representations are not valid in the whole
range down to the bound. As long as we are on the hyperboloid, related to
the original $SO(2,N)$, due to time periodicity $\alpha$ has to be integer.
On the universal cover, i.e. on $AdS_{N+1}$ $\,\alpha$ is a continuous
parameter.
\section{Propagators}
We now are interested in the quantum mechanical propagation kernel
$\langle X'|X\rangle$, where $|X\rangle$ is understood as a state
associated to the particle localized around the point $X$ on the
hyperboloid $\hh^{R}_{N} $. The states $|X\rangle $ have to
transform under a $SO(2,N)$ transformation $X\mapsto \Lambda X$,
with unitary representative $ \hat U_\Lambda$, as $|X\rangle
\mapsto \hat U_\Lambda|X\rangle =|\Lambda X\rangle $. For the
transition from the $\zeta$-representation ($\psi(\zeta )=\langle
\zeta|\psi \rangle$) of the previous section we need
$\langle\zeta|X\rangle$. Analyzing with (\ref{repr}) the equations
arising from
\begin{equation}\label{U}
 \langle\zeta|\,\hat U_\Lambda\,|X\rangle=
 \langle\, \hat U_{\Lambda^{-1}} \zeta|X\rangle ~,
\end{equation}
one gets up to a normalization factor
\begin{equation}\label{zeta|X}
\langle\zeta|X\rangle = [X_0-iX_{0'}+2\zeta_n\,X_n+
\zeta^2\,(X_0+iX_{0'})]^{-\alpha}.
\end{equation}
One can check that (\ref{zeta|X}) solves the Klein-Gordon equation
with mass squared equal to $R^{-2}\alpha(\alpha -N)$, which is in
consistence with the group theoretical interpretation given at the
end of the last section.  

Then the propagation kernel $\langle X'|X\rangle$ using (\ref{psi|psi})
is represented by
\begin{equation}\label{X|Y}
\langle X'|X\rangle= \int d^{N}\zeta\,
d^{N}\zeta^*~\frac{(1-2\zeta^*\zeta+{\zeta^*}^2\zeta^2)^{\alpha
-N}}{(Z'+2\zeta_n^*X_n'+Z'^*\zeta ^{*2})^{\alpha}~  (Z^*+2\zeta_nX_n+Z\zeta
^2)^{\alpha}}~,
\end{equation}
with $Z=X_0+iX_{0'}$. The integral is divergent for coinciding ($X'=X$)
or antipodal ($X'=-X$) points. As a natural regularization, respecting
the symmetry under \\
$(Z,X_n)\leftrightarrow (Z'^*,X'_n)$, we choose
$\theta \rightarrow \theta +i\epsilon ,~\theta '\rightarrow \theta
'-i\epsilon$. 

From the representation (\ref{X|Y}) one finds the
following interesting recursion in $\alpha$
\begin{equation}\label{rec}
\langle X'|X\rangle _{\alpha +1}=\frac{1}{2\alpha ^2}~(2\partial
^2_{Z'Z^*}-\partial ^2_{X_nX'_n}+2\partial ^2_{ZZ'^*})~\langle
X'|X\rangle _{\alpha }~.
\end{equation}
The evaluation of the integral in (\ref{X|Y}) becomes most
simply for $N=1$ and $\alpha =1$
\begin{equation}\label{11}
\langle X'|X\rangle _{1}=\frac{1}{4R^2}\log
\frac{(X-X')^2-i\epsilon\sin(\theta '-\theta)-4R^2}{(X-X')^2-i\epsilon\sin(\theta '-\theta) }~.
\end{equation}
The field theoretical Feynman
propagator $G_F(X,X')$ is given as usual by
\begin{equation}\label{Feyn}
G_F(X',X)=\Theta (T_{X'X})~\langle X'|X\rangle ~+~\Theta (T_{XX'})~\langle
X|X'\rangle~.
\end{equation}
Here the argument of the step function $\Theta$ is
$T_{X'X}=\sin(\theta '-\theta)$ for $\hh^{R}_{N}$ ($SO(2,N)$
invariant time ordering), or $T_{X'X}=(x_0'-x_0)$ for its
universal cover, i.e. $AdS_{N+1}$. The resulting propagator indeed
coincides with the well-known field theoretical expression, see
e.g. \cite{HF}. The precise handling of the distributional properties
of (\ref{11}) and (\ref{Feyn}) generated by our regularization of the
integral in (\ref{X|Y}) agrees with \cite{dulle}.
\section{Conclusions}
Based on the dynamics of massive spinless particles in $AdS$
space-time we constructed explicitly UIR's of $SO(2,N)$ for
generic $N$ acting on spaces of holomorphic functions of $N$
variables. The construction followed the lines of geometric
quantization. In agreement with general $SO(2,N)$ representation
theory these representations are labelled by the lowest energy
value $\alpha$. They are valid for $\alpha >N-1$ for $N\geq 2$ or
$\alpha >1/2$ for $N=1$. These bounds are above the well-known
unitarity bound $\alpha >N/2-1$. It remains to find a physical
interpretation why these representations do not exhaust the whole
$\alpha $ range allowed by unitarity.

Meanwhile we succeeded in constructing representations realized in terms of
canonical variables. They are well defined for all $\alpha $ allowed by
unitarity. A corresponding publication \cite{DJ}, in addition, will contain
a treatment of the special situation of enhanced symmetry at $\alpha =N/2\pm
1/2$, which gives the quantized particle a mass just enabling Weyl invariance
in the associated field theoretic action.

Part of our motivation was to explore the potential of particle quantization
techniques for the construction of quantum field propagators. Although their
explicit form for scalar fields on $AdS$ is well-known for all masses and
dimensions, for $AdS\times S$ only in special cases explicit closed
expressions are available \cite{DSS}.  There remains no doubt
that propagators can be constructed this way. However, our study in section 4
apparently did not show up this  method as more efficient than
the direct investigation of the field theoretical differential
equations. The potential of the recursion formula still has to be explored.
\\[3mm]
{\bf Acknowledgment:} We thank H.A. Kastrup, H. Nicolai, C. Sieg, M. Salizzoni
and A. Torrielli
for discussions. This research was supported by grants from DFG, INTAS,
RFBR and GAS.

\end{document}